\documentclass[pra,groupedaddress,superscriptaddress,onecolumn,amsmath,amssymb,a4paper,preprint,notitlepage]{revtex4}
\usepackage{geometry}
\usepackage[dvipsnames]{xcolor}
\usepackage{graphicx}
\usepackage{verbatim}
\usepackage{float}
\usepackage{graphics}
\usepackage{mathrsfs}
\usepackage{amssymb}
\usepackage{amsmath}
\usepackage{amsthm}
\usepackage{bm}
\usepackage{dsfont}
\usepackage{hyperref}
\usepackage{braket}
\usepackage{wrapfig}
\usepackage{amsbsy}
\usepackage{color}
\usepackage[normalem]{ulem}
\usepackage{subfig}

\usepackage[T1]{fontenc}

\newcommand{\be}{\begin{equation}}
\newcommand{\ee}{\end{equation}}
\newcommand{\ba}{\begin{eqnarray}}
\newcommand{\ea}{\end{eqnarray}}
\newcommand{\bibt}{\bibitem}

\begin{document}

\title{
Two-qubit entanglement generation through non-Hermitian Hamiltonians induced by repeated measurements on an ancilla
}

\today

\author{R. Grimaudo}
\affiliation{Dipartimento di Fisica e Chimica dell'Universit\`a di
Palermo, Via Archirafi, 36, I-90123 Palermo, Italy}

\author{A. Messina}
\affiliation{Dipartimento di
Matematica ed Informatica dell'Universit\`a di Palermo, Via
Archirafi, 34, I-90123 Palermo, Italy}

\author{A. Sergi}
\affiliation{Istituto Nazionale di Fisica Nucleare, Sez. di
Catania, Catania 95123, Italy}
\affiliation{Universit\`{a} degli
Studi di Messina, Dipartimento di Scienze Matematiche e
Informatiche, Scienze Fisiche e Scienze della Terra, viale F.
Stagno d'Alcontres 31, 98166 Messina, Italy}
\affiliation{Institute
of Systems Science, Durban University of Technology, P. O. Box
1334, Durban 4000, South Africa}

\author{N. V. Vitanov}
\affiliation{Department of Physics, St. Kliment Ohridski University of Sofia, 5 James Bourchier Boulevard, 1164 Sofia, Bulgaria}

\author{S. N. Filippov}
\affiliation{Steklov Mathematical Institute of Russian Academy of
Sciences, Gubkina St. 8, Moscow 119991, Russia}

\begin{abstract}
In contrast to classical systems, actual implementation of non-Hermitian Hamiltonian dynamics for quantum systems is a challenge because the processes of energy gain and dissipation are based on the underlying Hermitian system-environment dynamics that is trace preserving.
Recently, a scheme for engineering non-Hermitian Hamiltonians as a result of repetitive measurements on an anicillary qubit has been proposed.
The induced conditional dynamics of the main system is described by the effective non-Hermitian Hamiltonian arisng from the procedure.
In this paper we demonstrate the effectiveness of such a protocol by applying it to physically relevant multi-spin models, showing that the effective non-Hermitian Hamiltonian drives the system to a
maximally entangled stationary state.
In addition, we report a new recipe to construct a physical scenario where the quantum dynamics of a physical system  represented by a given non-Hermitian Hamiltonian model may be simulated.
The physical implications and the broad scope potential applications of such a scheme are highlighted.
\end{abstract}

\maketitle

\section{Introduction}

Historically, Gamow~\cite{gamow} was the first to adopt a
non-Hermitian Hamiltonian in order to study the radiative decay of
nuclei. There are also a number of other instances where
non-Hermitian Hamiltonians are useful~\cite{moiseyev}. For
example, this happens when one wants to study the parity-time (PT)
symmetry properties of the Hamiltonian~\cite{bender-pt}. Another
theory formulated in terms of non-Hermitian Hamiltonians is
obtained through the introduction of complex scaling
transformations~\cite{moiseyev}. Effective non-Hermitian
Hamiltonians are also obtained when, from a space comprising
discrete and continuous states, the continuous states are
projected out~\cite{cohen,f-1958,f-1962}. The Fock and Krylov
theorem~\cite{fock-krylov} states that the necessary and
sufficient condition for the presence of true decaying states is
that there must be a continuous part of the spectrum. Hence, the
projection operator formalism~\cite{cohen,f-1958,f-1962} and
non-Hermitian Hamiltonians provide an effective way to describe
decaying states. Non-Hermitian Hamiltonians can also be postulated
on the basis of physical
considerations~\cite{as-kz-2013,kz-as-2014,kz-2015,kz-2016,kz-2017},
in order to describe gain or loss of probability. Dynamics in
terms of non-Hermitian Hamiltonians has been investigated for
quantum~\cite{lp-kz-2018} and quantum-classical
systems~\cite{as-2015}, adopting the phase space representation
quantum mechanics. The dynamics of non-Hermitian quantum
mechanical systems can be studied either in terms of a linear
equation for a non-normalized density
matrix~\cite{as-kz-2013,grimaudo-2018} or in terms of a non-linear
equation for a normalized density
matrix~\cite{as-kz-2013,grimaudo-2018}. Upon combining the linear
evolution for the non-normalized density matrix and the non-linear
equation of motion for the normalized density matrix, different
forms of correlations functions~\cite{as-kz-2015} and
entropies~\cite{as-kz-2016,as-pvg-2016} have been defined.

Despite such developments in the theoretical realm, so far
the observation of non-Hermitian dynamics in experimental
situations has been somewhat limited to classical
dissipative systems whose theoretical description was
mapped onto that provided by quantum-like non-Hermitian (and
often PT-symmetric) Hamiltonians.
Examples of such systems are given by
optical lattices~\cite{makris-2008,zyablovsky-2014},
optical radiation interacting
with atomic systems~\cite{hang-2013,peng-2016,zhang-2016},
electronic circuits~\cite{bender-2013,assaw-2017,choi-2018},
microwave billiards~\cite{bittner-2012}
simple mechanical systems~\cite{cbender-2013},
and acoustical systems~\cite{zhu-2014,popa-2014,fleury-2015}.
In all these cases, the Non-Hermitian dynamics of classical
systems is well understood and experimentally realized by means
of asymmetric attenuation and amplification.
However, the experimental realization of true non-Hermitian
quantum systems (\emph{i.e.}, which do not arise from an
isomorphism between classical dissipative dynamics and
non-Hermitian quantum mechanics) is difficult since quantum
systems naturally obey the laws of Hermitian quantum mechanics.
For example, both the attenuation and amplification of signals are
described by physical quantum channels (completely positive and
trace preserving maps) with Hermitian Hamiltonian dynamics involving
the system and its environment~\cite{holevo,filippov-jms-2019,filippov-2014}.

In order to demonstrate the occurrence of non-Hermitian Hamiltonians, some theoretical methods have been
proposed. Among these, we highlight those based on the universal
concept of dilation mapping~\cite{z-2020} of a non-Hermitian Hamiltonian
into a Hermitian Hamiltonian living in a higher dimensional Hilbert
space~\cite{gunther-2008,kawabata-2017,teretenkov-2019,huang-2019,wu-2019}.
Interestingly, the dilation mapping is, broadly speaking, the inverse of the projection formalism~\cite{cohen,f-1958,f-1962},
according to which one projects a Hermitian Hamiltonian into a non-Hermitian Hamiltonian,
defined in a lower dimensional Hilbert space.
Although all the known schemes exploit the general
concept of dilation/inverse-projection formalism in order
to propose experimental schemes for building non-Hermitian Hamiltonians, the actual
implementation of these schemes is tailored in some way to a
chosen specific system. For instance, the authors of
Ref.~\cite{wu-2019} use a time-dependent Hermitian Hamiltonian in
a higher dimensional Hilbert space of two qubits in order to
simulate a non-Hermitian Hamiltonian for a single qubit.

Quite recently, an experimental scheme implementing the quantum
dynamics of a finite-dimensional system $S$ generated by a
non-Hermitian Hamiltonian operator, has been reported
\cite{luchnikov-2017}. The basic idea is to couple $S$ with a
quantum ancilla subsystem $A$ and to follow the time evolution of
$S$ conditioned by a Zeno measurement protocol applied  on the
ancilla finite-dimensional subsystem only. In accordance with the
previosly quoted Ref. \cite{luchnikov-2017}, the reduced density
matrix of $S$, conditioned by the progression of collapses induced
in this way on the  state of the combined system $S+A$, evolves
under the action of an effective non-Hermitian Hamiltonian which
may be explicitly constructed in the so called stroboscopic limit.
We observe that the proposal of Ref. \cite{luchnikov-2017} differs
from that of Feshbach since the latter is not a conditional one
and since the ancilla subsystem can hardly be considered
dynamically equivalent to an environment with infinite degrees of
freedom, as requested in the projection method invented by
Feshbach. In addition, the scheme proposed in Ref.
\cite{luchnikov-2017} can be easily experimentally implemented
and, form a theoretical point of view, leads to a solvable quantum
dynamical problem.

Considering the idea of Ref.~\cite{luchnikov-2017}, it is interesting
to note the following.
Continuous measurements on the ancilla generate infinitesimal lifetimes
for its states.
Hence, the time-energy uncertainty principle makes sure that the
energy of the ancilla under continuous measurements cannot be sharply peaked.
It follows that the ancilla under continuous
measurements effectively acts as a continuum of states with which the system
interacts. A similar reasoning is found in Ref.~\cite{dattoli}.
According to the theorem of Fock and Krylov~\cite{fock-krylov},
once the system is in contact with a continuum of states, provided by the
ancilla under continuous measurements, it evolves experiencing the decay of its
states. The limited lifetime of decaying states~\cite{fonda-ghirardi-rimini}
and the representation of the width of the energy levels
by means of an imaginary component of the system's eigenvalues
naturally lead to a non-Hermitian Hamiltonian.

The first goal of this work is to prove theoretically that entanglement
in a system of two interacting qubits can be generated by means of
stroboscopic measurements on a third qubit, coupled to the first
two, for which it constitutes the ancilla subsystem $A$ (requested
by Ref.~\cite{luchnikov-2017}). Continuous measurements on the
ancilla, that is a Zeno measurement protocol, produce an effective
non-Hermitian Hamiltonian determining the time evolution of the
reduced and conditioned density matrix of the two spin system $S$.
Two main results must be emphasized: 1) the possibility of generating maximally entangled states of the two qubits thanks to the repeated measurements on the ancilla; 2) the possibility of getting information about the (an)isotropy level of the pair-wise interactions between the three qubits, by studying the effective dynamics of the two-qubit system.

The experimental protocol reported in Ref.~\cite{luchnikov-2017} (hereafter referred to as direct) realizes a dynamical constraint under which the system $S$ is effectively driven in its Hilbert space as if it were subjected to a non-Hermitian Hamiltonian model.
In this paper, we successfully face with the following inverse problem: given a non-Hermitian Hamiltonian model at will, describing the quantum dynamics of a physical system $S$, to find a Hermitian model reproducing the assigned non-Hermitian Hamiltonian of the direct procedure presented in Ref.~\cite{luchnikov-2017}.
In principle, solving this inverse problem means associating to an arbitrary non-Hermitian model a physical scenario where its quantum dynamics can be experimentally simulated.

This manuscript is organized in the following way. In Sec.~\ref{sec:nhrho} the formalism describing the time evolution of the reduced density matrix when the Hamiltonian of the relevant system is non-Hermitian is outlined.
In Secs.~\ref{sec:filippov} we discuss at length the general direct scheme for experimentally realizing  a (\textit{a priori} unknown) non-Hermitian Hamiltonian.
In Sec.~\ref{sec:inverse} the recipe for solving the inverse problem is reported.
The application of the direct protocol to a (two+one)-qubit system ($S+A$) model and the detailed study of the effective non-Hermitian dynamics of the resulting two spin-qubit system is developed in Sec. \ref{Ent}, where remarkable physical effects, which are suitable for experimental and technological applications, are brought to light.
Finally, conclusive remarks and comments are discussed in the last
section.

\section{Density matrices and effective non-Hermitian Hamiltonians}
\label{sec:nhrho}

Let us assume that the dynamics of a quantum system $S$, living in a discrete Hilbert space,
is described by a non-Hermitian Hamiltonian, $H_{\rm eff}\neq H_{\rm eff}^\dag$.
If $S$ is appropriately coupled to an environment living in a continuous Hillbert space,
the projection operator formalism~\cite{cohen,f-1958,f-1962} allows one to derive such a non-Hermitian
Hamiltonian describing the system~\cite{faisal-moloney}.
Then, in terms of the non-Hermitian Hamiltonian~\cite{baker} $H_{\rm eff}$,
the Schr\"odinger equation reads
\be
\frac{d}{d t} \ket{\Psi(t)} = -iH_{\rm eff}\ket{\Psi(t)}
\;.\label{eq:Psi}
\ee
In Eq.~(\ref{eq:Psi}) and in the following we assume units of measure
such that $\hbar=1$.
In accordance with the theorem of Fock and Krylov~\cite{fock-krylov},
stating that the necessary and sufficient condition for the existence
of true decaying states for $S$ is the interaction with a continuum of states,
Eq.~(\ref{eq:Psi}) describes the decay of the system's states.
As a matter of fact, each solution $\ket{\Psi(t)}$, solution of a Cauchy problem
for the Eq.~(\ref{eq:Psi}) can also be written in term
of a non-unitary propagator $U(t)\neq U^\dag(t)$~\cite{dattoli,baker}:
\be
|\Psi(t)\rangle=U(t)|\Psi(0)\rangle\;,
\label{eq:upsi}
\ee
which clearly shows that the probability for the system is not conserved.
Upon defining a non-normalized density matrix as
$\rho=|\Psi(t)\rangle\langle\Psi(t)|$, one can easily derive its equation of motion:
\be
\frac{d}{d t} \rho(t)=
-\frac{i}{\hbar}\left( H_{\rm eff}\rho(t) - \rho(t) H_{\rm eff}^\dag\right)\;,
\label{eq:dOmedt-quasi-comm}
\ee
whose solution can be written $U(t)\rho(0)U^\dag(t)$, as usual.
When $H_{\rm eff}$ is time-independent (as we will assume),
the non-unitary propagator introduced in Eq.~(\ref{eq:upsi}),
can be written as $U(t)=\exp(-iH_{\rm eff}t)$.

The non-Hermitian Hamiltonian $H_{\rm eff}$ can always be defined in
terms of the sum of a Hermitian Hamiltonian,
$H_0=(H_{\rm eff} +H_{\rm eff}^\dag)/2$
and an anti-Hermitian operator,
$i\Gamma=-(H_{\rm eff}-H_{\rm eff}^\dag)/2$.
The Hermitian operator $\Gamma$ is called the decay operator.
Combining Eq.~(\ref{eq:Psi}) with its adjoint~\cite{graefe-schubert},
one obtains the equation of motion~\cite{as-kz-2013}
for the density matrix introduced right above Eq.~(\ref{eq:dOmedt-quasi-comm})
\ba
\frac{d}{d t}\rho(t)= -\frac{i}{\hbar}\left[H_0,\rho(t)\right]
-\frac{1}{\hbar} \left\{\Gamma,\rho(t)\right\} \;,
\label{eq:dOmedt-antic}
\ea
where $[\cdot,\cdot]$ is the commutator
and $\{\cdot,\cdot\}$ is the anti-commutator.
Equations~(\ref{eq:dOmedt-quasi-comm}) and~(\ref{eq:dOmedt-antic})
reduce to the standard ones when $\Gamma=0$ (which means that
$H_{\rm eff}=H_0$ is Hermitian).
Eqs.~(\ref{eq:dOmedt-quasi-comm}) and~(\ref{eq:dOmedt-antic})  keep their valididy even if
the system is initially not in a pure state $|\Psi(0)\rangle$
but in a mixture of states $\ket{\Psi^{(i)}(0)}$.

Equations~(\ref{eq:dOmedt-quasi-comm}) and~(\ref{eq:dOmedt-antic})
do not conserve the trace of the non-normalized density matrix
$\rho(t)$. Upon taking the trace of Eq.~(\ref{eq:dOmedt-antic}),
one gets \ba \frac{d}{d t} {\rm Tr}\left[\rho(t)\right]=
-\frac{2}{\hbar} {\rm Tr}\left[\Gamma\rho(t)\right] \;.
\label{eq:dOTrdt} \ea Equation~(\ref{eq:dOTrdt}) shows explicitly
that the probability of the system is not conserved. Upon
normalizing $\rho(t)$ at every $t$ with its time-dependent trace,
one can define a normalized density matrix $\varrho(t)$ given by
\ba \varrho(t)=\frac{\rho(t)}{{\rm Tr}\left[\rho(t)\right]}
=\frac{e^{-i H_{\rm eff} t}\rho(0)e^{i H_{\rm eff}^\dag t}} {{\rm
Tr}\left[e^{-i H_{\rm eff} t}\rho(0)e^{i H_{\rm eff}^\dag t}
\right]} =\frac{U(t)\rho(0)U^\dag(t)} {{\rm
Tr}\left[U(t)\rho(0)U^\dag(t) \right]} \;. \ea The equation of
motion obeyed by $\varrho(t)$
is~\cite{as-kz-2013,kz-2015,as-kz-2015,as-kz-2016,grimaudo-2018}
\be \frac{d}{d t} \varrho(t)=
-\frac{i}{\hbar}\left[H_0,\varrho(t)\right]-\frac{1}{\hbar}
\left\{\Gamma,\varrho(t)\right\} +\frac{2}{\hbar}{\rm
Tr}\left[\Gamma \varrho(t)\right] \varrho(t) \;. \label{eq:drhodt}
\ee
Equation~(\ref{eq:drhodt}) is a non-linear equation that, by construction,
preserves the trace of $\varrho(t)$.
Averages of dynamical variables, which are represented by operators denoted
here with $\chi$, are calculated in the the standard way
\be
\langle \chi(t)\rangle \equiv {\rm Tr}\left[\varrho(t)\chi\right] \;.
\label{eq:ave}
\ee
When the operator $\chi$ is Hermitian, the average in Eq.~(\ref{eq:ave}) is
real.
Equation~(\ref{eq:drhodt}) reduces to the standard
linear equation of Hermitian quantum mechanics when $\Gamma=0$.

Equations~(\ref{eq:Psi})--(\ref{eq:ave}) imply that,
notwithstanding the non-Hermitian Hamiltonian $H_{\rm eff}$, a
non-linear Hermitian quantum mechanics can be defined. In the next
Section, it is shown that the structure of the nonlinear equation
in~(\ref{eq:drhodt}) turns out to be useful in a different context
too, where the system is coupled to a continuously measured
ancilla.

\section{Non-Hermitian Hamiltonians due to repeated measurements}
\label{sec:filippov}

\begin{figure}
\centering
\includegraphics[width=6cm]{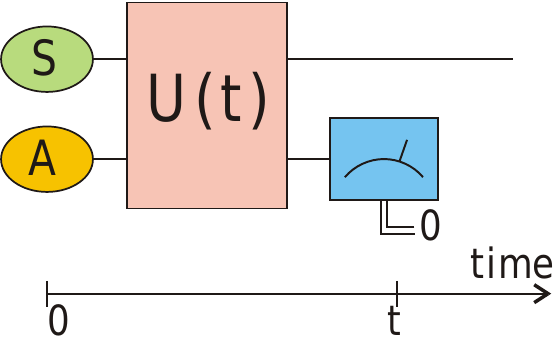}
\caption{Conditional implementation of non-unitary dynamics for
system $S$ via projective measurement on ancilla $A$.}
\label{figure-1}
\end{figure}

When one studies a bipartite system undergoing unitary quantum
dynamics, there is a possibility to physically implement
\textit{conditionally} a non-unitary dynamics for one of its
subsystems, with the resulting effective Hamiltonian being
non-Hermitian~\cite{wu-2019,luchnikov-2017,wen-2019}. The idea is
to let a quantum system $S$ interact with an ancillary quantum
system $A$ for some time $t$ and then perform a projective
measurement on the ancilla, see Fig.~\ref{figure-1}.
The Hamiltonian operator $H$, describing such a coupling, is supposed time-independent and experimentally implementable.
Provided the ancilla is originally prepared in the pure non-degenerate state
$\ket{0_A}$ and measured in the orthonormal basis
$\{\ket{0_A},\ldots\}$, with the result being $0$, the reduced
density operator of the system $S$ at the time $t$ collapses into
the following unnormalized conditional density matrix:
\begin{equation} \label{transformation-general}
\begin{aligned}
\rho_S(0) \longrightarrow
\rho_S^c(t) &= \bra{0_A} \,\, U(t)
\,\, \left[ \rho_S(0) \otimes \ket{0_A}\bra{0_A} \right] \,\,
U^{\dag}(t) \,\, \ket{0_A} \\
&= [\bra{0_A} \,\, U(t) \ket{0_A}]
\,\, \bra{0_A }\left[ \rho_S(0) \otimes \ket{0_A}\bra{0_A} \right] \ket{0_A} \,\,
[\bra{0_A} U^{\dag}(t) \,\, \ket{0_A} ] \\
&\equiv K(t) \rho_S(0) K^{\dag}(t),
\end{aligned}
\end{equation}

\noindent where in the first equality we introduced the identity
operator $I_A=\sum_j\ket{j_A}\bra{j_A}$, while in the last passage
we defined in the Hilbert space of the system $S$ the operator
$K(t) \equiv  \bra{0_A} \, U(t) \, \ket{0_A}$. $\rho_S(0)$
represents the initial reduced density matrix of the subsystem
$S$, while $\rho_S^c(t)$ stands for the non-normalized reduced
conditional density matrix stemming from the measurement act
performed on the ancilla qubit at $t > 0$. $U(t)$ is the unitary
evolution operator for the whole system $S+A$ governed by $H$, while $K(t)$, as a
submatrix of a unitary matrix, is a non-unitary evolution operator
for the system $S$. Thus, $\rho_S^c(t)$ is not a density matrix
since its evolution does not preserve its trace. The success
probability to observe the outcome $0$ while measuring the ancilla
qubit at the time $t$ is
\begin{equation}\label{success probability}
p_{\rho_S(0)}(t) = \bra{0_A} \rho_A(t) \ket{0_A} = \bra{0_A} \,\,
{\rm tr}_S \left\{ U(t) \,\, \left[ \rho_S(0) \otimes
\ket{0_A}\bra{0_A} \right] \,\, U^{\dag}(t) \right\} \,\,
\ket{0_A} = {\rm tr}_S \rho_S^c(t) \equiv {\rm tr} \rho_S^c(t),
\end{equation}

\noindent and depends on the initial system state $\rho_S(0)$. The
trace of the subnormalized operator $\rho_S^c(t)$ determines how
often the desired event takes place. The properly normalized
density operator reads $\varrho_S^c(t) = \rho_S^c(t) / {\rm tr}
\rho_S^c(t) = \rho_S^c(t) / p_0(t)$.


It is worthwhile to point out that the authors of Ref.~\cite{wu-2019}
simulate the qubit evolution with a time-independent non-Hermitian
Hamiltonian $H_{\rm eff}$ by controlling the actual evolution
operator $U(t)$ for the system and ancilla.
In this case, the
Hamiltonian for the system and ancilla is time-dependent because
$U(t)$ is not a semigroup, which requires a sophisticated driving.
Repeated measurements on ancilla help overcome this drawback in
the stroboscopic limit~\cite{luchnikov-2017}, when the dynamics of
ancilla is effectively frozen.

\begin{figure}
\centering
\includegraphics[width=12cm]{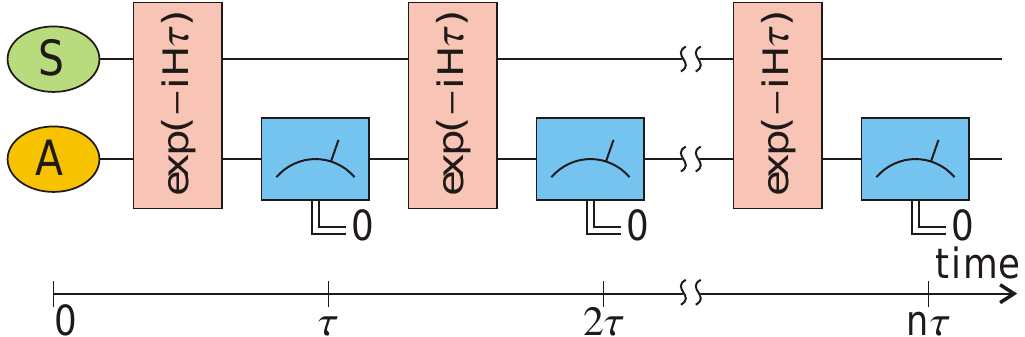}
\caption{Repeated measurements on ancilla $A$ result in
non-Hermitian Hamiltonian dynamics for system $S$.}
\label{figure-2}
\end{figure}

Suppose the ancilla is initially in the non-degenerate state
$\ket{0_A}$ and is repeatedly measured after equal time intervals
$\tau$ in the basis $\{\ket{0_A}, \ldots\}$. Considering the
physical meaning of Eq. \eqref{transformation-general},
$\rho_S^c(\tau)$ determines the new initial state of $S$ which,
tensorially multiplied by $\ket{0_A}\bra{0_A}$, gives the new
(non-normalized) initial condition for the total system $S+A$
after the measurement act at the time instant $\tau$. Provided $n$
sequential measurements give the outcome 0, the system state, in
view of Eq. \eqref{transformation-general}, collapses into
\begin{equation}\label{rho s c n tau}
\rho_S(0) \longrightarrow \rho_S^c(n\tau) = K^n(\tau)
\rho_S(0) \left( K^n(\tau) \right)^{\dag}.
\end{equation}

\noindent The probability $p(n\tau)$ for observing $n$ sequential
outcomes $0$ while measuring the ancilla qubit $n$ times is
\begin{eqnarray}
p(n\tau) &=& p_{\rho_S(0)}(\tau) \times
p_{\varrho_S^c(\tau)}(\tau)
\times \ldots \times p_{\varrho_S^c((n-1)\tau)}(\tau) \nonumber\\
&=& {\rm tr}[K(\tau) \rho_S(0) K^{\dag}(\tau)] \times {\rm
tr}[K(\tau) \varrho_S^c(\tau) K^{\dag}(\tau)] \times \ldots \times
{\rm
tr}[K(\tau) \varrho_S^c((n-1)\tau) K^{\dag}(\tau)] \nonumber\\
&=& {\rm tr}[K(\tau) \rho_S(0) K^{\dag}(\tau)] \times \frac{{\rm
tr}[K(\tau) \rho_S^c(\tau) K^{\dag}(\tau)]}{{\rm tr}[K(\tau)
\rho_S(0) K^{\dag}(\tau)]} \times \ldots \times \frac{{\rm
tr}[K(\tau) \rho_S^c((n-1)\tau) K^{\dag}(\tau)]}{{\rm tr}[K(\tau)
\rho_S^c((n-2)\tau) K^{\dag}(\tau)]} \nonumber\\
&=& {\rm tr}[K(\tau) \rho_S^c((n-1)\tau) K^{\dag}(\tau)] = {\rm
tr} [K^n(\tau) \rho_S(0) \left( K^n(\tau) \right)^{\dag}] = {\rm
tr}\rho_S^c(n\tau), \label{prob-n-tau}
\end{eqnarray}

\noindent which means that $p(n\tau)$ is merely the trace of the
unnormalized operator $\rho_S^c(n\tau)$ and monotonously
diminishes.

Since the Hamiltonian of the combined closed system $S+A$ is time
independent, the operator $K(\tau) = \bra{0_A} \exp(-i H \tau)
\ket{0_A}$ is time-independent as well. The Taylor expansion with
respect to $\tau$ yields
\begin{eqnarray} \label{K-tau}
K(\tau) &=& \bra{0_A} \left(I_{S+A} - i \tau H -
\frac{\tau^2}{2} H^2 + o_{S+A}(\tau^2) \right) \ket{0_A} \nonumber\\
&=& I_S - i \tau \bra{0_A} H \ket{0_A} - \frac{\tau^2}{2}
\bra{0_A} H^2 \ket{0_A} + o_S(\tau^2) \nonumber\\
&=& \exp\left( - i\tau H_0^{\rm S} - \frac{\tau^2}{2} \Gamma_{\rm
S} + o_S(\tau^2) \right),
\end{eqnarray}

\noindent where $o(\tau^2)$ denotes an operator acting on the
corresponding Hilbert space ($S+A$ or $S$) such that its norm
$\|o(\tau^2)\|$ satisfies $\lim_{\tau \rightarrow 0} \|o(\tau^2)\| /
\tau^2 = 0$,
\begin{equation} \label{H1-H2}
H_0^{\rm S} = (H_0^{\rm S})^{\dag} =  \bra{0_A}  H \ket{0_A} \quad \text{and} \quad
\Gamma_{\rm S}=\Gamma_{\rm S}^{\dag}=\bra{0_A}H^2\ket{0_A}-(H_0^{\rm S})^2
= \bra{0_A} H
\ket{1_A} \, \bra{1_A} H  \ket{0_A} \geq 0.
\end{equation}

If the measurement repetition rate $\frac{1}{\tau}$ is much
greater than the maximum Bohr frequency of $H$, then one can
neglect the term $o_S(\tau^2)$ in Eq.~\eqref{K-tau}. This means
that the system evolution in between the measurement acts is
infinitesimal, so that the stroboscopic time $t = n\tau$ is
quasi-continuous in full analogy with the quantum collision
models~\cite{rau-1963,giovannetti-2012,filippov-2017,fsp-2020}. As
a result, we obtain
\begin{equation}\label{Kton}
K^n(\tau) = \exp\left( - i n\tau H_0^{\rm S}  - \frac{n \tau^2}{2} \Gamma_{\rm S} +
o(\tau^2) \right) = \exp\left[ - i t \left( H_0^{\rm S}  - \frac{i
\tau}{2} \Gamma_{\rm S} \right) + o(\tau^2) \right]
\end{equation}

\noindent and the resulting non-Hermitian Hamiltonian, effectively
acting on $S$ under the Zeno experimental protocol for repeated
measurements on the ancilla, may be taken as
\begin{equation} \label{H-eff}
H_{\rm eff} = H_0^{\rm S}  - \frac{i \tau}{2} \Gamma_{\rm S}.
\end{equation}

At a timescale much greater than $\tau$, the dynamics of the
subnormalized density operator $\rho_S^c(t)$ is quasi-continuous
and, in view of Eq. \eqref{Kton}, it satisfies the equation
\begin{equation}
\frac{d\rho_S^c(t)}{dt} = -i \left( H_{\rm eff} \rho_S^c(t) -
\rho_S^c(t) H_{\rm eff}^{\dag} \right).
\end{equation}

Eq.~\eqref{prob-n-tau} implies that the quasi-continuous
probability $p(t)$ of successful observation of desired
measurement outcomes (all 0s) up to time $t$ diminishes in time in
accordance with the equation
\begin{equation}
\frac{dp(t)}{dt} = \frac{d \, {\rm tr}[\rho_S^c(t)]}{dt} = - \tau
{\rm tr} [ \Gamma_{\rm S} \rho_S^c(t) ],
\end{equation}

\noindent with $\frac{dp(t)}{dt} \leq 0$ because $\Gamma_{\rm S}$
is positive semidefinite.

As commented before, this circumstance is due to the
non-Hermiticity of $H_{\text{eff}}$ and the consequent effective
non-unitary time-evolution operator $K(t)$ for the two-qubit
system. It means that we cannot exploit $\rho_S^c(t)$ to get
relevant statistically valid information about the two-qubit
system. In order to have a physically admissible density operator,
we simply normalize the reduced, conditional density operator as
follows
\begin{equation} \label{normalized}
\varrho_S^c(t)=\frac{\rho_S^c(t)}{{\rm tr}[\rho_S^c(t)]} = \frac{K^n(t)
\rho_S(0) (K^n)^{\dag}(t)}{{\rm tr}[K^n(t) \rho_S(0) (K^n)^{\dag}(t)]}
= \frac{ e^{-iH_{\rm eff}t} \rho_S(0) e^{+iH_{\rm eff}^{\dag}t}
}{{\rm tr}[ e^{-iH_{\rm eff}t} \rho_S(0) e^{+iH_{\rm
eff}^{\dag}t} ]}
\end{equation}
which, as we know from Sec.~\ref{sec:nhrho}, satisfies the following non-linear evolution equation:
\begin{equation}
\frac{d\varrho_S^c(t)}{dt} = -i [H_0^{\rm S},
\varrho_S^c(t)] - \frac{\tau}{2} \{\Gamma_{\rm S},
\varrho_S^c(t)\} + \tau {\rm tr}[\Gamma_{\rm S}
\varrho_S^c(t)] \varrho_S^c(t).
\end{equation}

\noindent It is important to underline that if $\rho_S(0) = \ket{\psi_S(0)}\bra{\psi_S(0)}$, then
the normalized density operator $\varrho_S^c(t)$ remains
pure during the evolution (has zero entropy) and the corresponding
wavefunction satisfies a non-linear equation
\begin{equation}
i \frac{d\ket{\psi_S(t)}}{dt} = \left( H_0^{\rm S} -  \frac{i\tau}{2} \Gamma_{\rm S}
\right) \ket{\psi_S(t)} + \frac{i \tau}{2} \bra{\psi_S(t)} \Gamma_{\rm S}
\ket{\psi_S(t)} \, \ket{\psi_S(t)}.
\end{equation}

The protocol leading to the non-Hermitian Hamiltonian given by Eq. \eqref{H-eff} assumes that the evolution of the combined system $S+A$ is governed by a time-independent Hamiltonian describing the coupling of $S$ with the ancilla subsystem $A$.
In view of the importance played by the time-dependent Hamiltonian models as control tools, it is worth examining where our protocol fails if the time-independence of $H$ is relaxed.
This analysis is of course useful  to understand the reasons of the restriction we introduce on $H$ and is necessary as well to highlight the possibility of extending this protocol to more general situations.

Let us begin by observing that Eqs. \eqref{transformation-general} and \eqref{success probability} are valid in both cases and in particular the formal introduction of the operator $K(t)$.
Equations \eqref{rho s c n tau} and \eqref{K-tau} are instead not valid.
In fact, if $H$ depends on time, in the interval $[\tau, 2\tau]$ the time evolution of $(S+A)$ is ruled out by a Hamiltonian different (due to its time dependence) from that generating the evolution of the system $S+A$ in the interval $[0,\tau]$.
This implies that $K^n (\tau)$ must be substituted by the product of $n$ generally different operators $K$-like always of argument $\tau$. Moreover, to find the analytical form of $U$ may be relatively more complicated as usual.

In practice, the approach required to generate the mathematical expression of $H_{eff}$ is not a trivial extension of the one reported in this paper and turns out to be more intricate.
However, the points we have elucidated somehow legitimate the expectation of arriving to a generalized protocol in the near future.
In particular, the analytical progress in this problem is achievable in the adiabatic regime, when, in addition to the time-dependent version of the stroboscopic approximation, one assumes that the characteristic frequency of classical fields (controlling Hamiltonian $H$) is much smaller than the measurement repetition rate $\tau^{-1}$.

\section{Non-Hermitian Hamiltonian engineering} \label{sec:inverse}

In Section~\ref{sec:filippov}, we considered a scenario in which
the experimentalist has a composite quantum system $S+A$, with some
fixed Hamiltonian $H$, on which he repetitively performs projective
measurements on the  qubit $A$ only. Exploiting this scheme  we then derived the effective
generally non-Hermitian Hamiltonian $H_{\rm eff}$ describing the quantum evolution of $S$.

In  this section we consider the following inverse problem. 
Suppose the Experimentalist is aimed at implementing the non-Hermitian Hamiltonian $H_{\rm eff}$  of the system S. Our  goal, from a theoretical point of view, is twofold. The first one is to provide the physical Hermitian Hamiltonian $H$ to be engineered in the lab for the enlarged system $S+A$, A being the qubit ancilla A coupled to S. The second one consists in prescribing as well the measurement repetition rate $\tau^{-1}$ under which the stroboscopic approximation  generates  the conditional reduced dynamics of $S$  as governed by the prescribed $H_{\rm eff}$ of interest. It is worthy emphasizing that, to achieve a wider applicability of the method we are going to describe,  from the very beginning we assume that $H_{\rm eff}$ acts on the Hilbert finite-dimensional space of S  where it is still representable as Gauss combination of two hermitian  operators mamely $H_1$ + i $H_2$. We underline that no assumption is here made concerning the spectra of $H_1$  and $H_2$ or whether $H_1$ + i $H_2$ is diagonalizable.

The resolution of the posed inverse problem proceeds as follows.
The Hermitian part of $H_{\rm eff}$ reads $\frac{1}{2}(H_{\rm eff} + H_{\rm eff}^{\dag})$ and corresponds to $H_0^{\rm S}$ in Eq.~\eqref{H1-H2}.
Calculate the Bohr frequencies for the Hermitian operator
$i(H_{\rm eff} - H_{\rm eff}^\dag)$ and denote by $f$ its maximum
Bohr frequency. Fix $\tau$ in such a way that $f \tau \ll 1$,
e.g., $\tau = 10^{-2} f^{-1}$.
Differently from the operator $\Gamma_{\rm S}$ in Eq.~\eqref{H1-H2}  that is
positive-semidefinite by construction, the operator $i(H_{\rm eff} - H_{\rm eff}^\dag)$, as previously claimed, does not possess, in general,  such a property.
Thus we introduce the  constant
$c=\max(0,-M)$, where $M$ is the minimum eigenvalue of the
operator $\frac{i}{\tau}(H_{\rm eff} - H_{\rm eff}^\dag)$. Note
that the dimensional parameter $c$ depends on the chosen measurement
repetition rate $\tau$ and that it would vanish if $i(H_{\rm eff} - H_{\rm eff}^\dag)$. Then the operator $c I +\frac{i}{\tau}(H_{\rm eff} - H_{\rm eff}^{\dag}) \geq 0$ and
corresponds to $\Gamma_{\rm S}$ in Eq.~\eqref{H1-H2}.
Finally, using the established correspondence and the explicit
formulae~\eqref{H1-H2}, we provide the total Hermitian Hamiltonian
for the system and the ancillary qubit
\begin{equation}\label{Inverse H}
H = \frac{1}{2}(H_{\rm eff} + H_{\rm eff}^{\dag}) \otimes
\ket{0_A}\bra{0_A} + \sqrt{c I + \frac{i}{\tau}(H_{\rm eff} -
H_{\rm eff}^{\dag})} \otimes \left( \ket{0_A} \bra{1_A} +
\ket{1_A}\bra{0_A} \right).
\end{equation}
By construction, the maximum Bohr frequency for $H$, which relates
the states $\ket{0}$ and $\ket{1}$ for the ancilla qubit $A$, is
of the order $\gamma = \sqrt{f / \tau }$, so it satisfies the
condition $\gamma \tau \ll 1$ if $f \tau \ll 1$. The latter
condition is satisfied as we prescribed the inter-measurement
duration time $\tau$ accordingly. This validates the stroboscopic
approximation.

The proposed scheme for engineering non-Hermitian Hamiltonians at will is
rather universal. Whatever finite-dimensional operator $H_{\rm eff}$ is given (no matter if it is either PT-symmetry or pseudo-Hermitian, no matter
if it is diagonalizable or not), there exists a Hermitian operator
$H$ of twice dimension (i.e., acting on the tensor product of the
original Hilbert space and a qubit ancilla) such that the reduced
dynamics of $S$ in the stroboscopic approximation is equivalent to the
quantum dynamics under the assigned non-Hermitian Hamiltonian $H_{\rm eff}$. Therefore,
formula~\eqref{Inverse H} explicitly prescribes a Hermitian
Hamiltonian for the whole system $S+A$ to simulate, with the help of the experimental
protocol described in Section~\ref{sec:filippov}, the reduced time evolution of $S$ generated by $H_{eff}$.
This last aspect is remarkable since such an inverse protocol is in particular applicable when $H_{eff}$ is pseudo-Hermitian. This means that Eq.~\eqref{Inverse H} provides a recipe to generate a Hermitian Hamiltonian and then, generally speaking, a physical scenario where the quantum dynamics of the pseudo-Hermitian Hamiltonian of interest may be simulated. In the class of non-Hermitian Hamiltonian models, pseudo-Hermiticity \cite{PseudoHermitian} occupies a special place since it is the most benign one, with various nice properties as, for example, the existence of invariants or the easy derivation of analytical solutions \cite{Simeonov,Torosov}.

\section{Non-Hermitian dynamics for coupled qubits induced by repeated measurements}\label{Ent}

\subsection{Symmetric two-qubit effective non-Hermitian Hamiltonian}

\begin{figure}
\centering
\includegraphics[width=6cm]{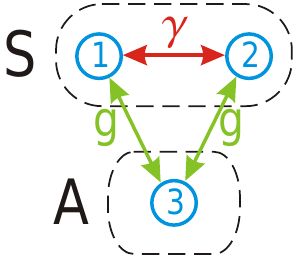}
\caption{Interaction graph for three spins coupled via
Hamiltonian~\eqref{three-spin-Hamiltonian}. First two spins
compose a bipartite system $S$ under study. Third spin is
auxiliary ($A$) and is subjected to repeated measurements.}
\label{figure-3}
\end{figure}

Consider three spin-$\frac{1}{2}$ particles (qubits) with the
pairwise interaction Hamiltonian (in units of $\hbar$, that is, $\hbar=1$)
\begin{eqnarray} \label{three-spin-Hamiltonian}
H &=& \gamma_{xy} ( \sigma_1^x \sigma_2^x + \sigma_1^y \sigma_2^y
) + \gamma_z \sigma_1^z \sigma_2^z + g_{xy} ( \sigma_1^x
\sigma_3^x + \sigma_1^y \sigma_3^y ) + g_z \sigma_1^z \sigma_3^z +
g_{xy} ( \sigma_2^x \sigma_3^x + \sigma_2^y \sigma_3^y ) + g_z
\sigma_2^z \sigma_3^z,
\end{eqnarray}

\noindent which generalizes so-called XXZ model~\cite{XXZ} and
assumes that the third auxiliary spin is equidistant from the
other two spins, see Fig.~\ref{figure-3}. Hereafter,
$\sigma^x,\sigma^y,\sigma^z$ denote the conventional set of Pauli
operators.

We consider the first two spins as a system ($S$) and the third
spin as an ancilla ($A$), whose spin projection onto $z$-axis is
repeatedly measured after equal time intervals $\tau$. If the
initial state of the third spin is $\ket{0_A}$ such that
$\sigma_3^z \ket{0_A} = \ket{0_A}$ and the measurements confirm
the spin remains in this state, then the system, in the stroboscopic limit, experiences a
non-unitary evolution with the effective non-Hermitian
Hamiltonian~\eqref{H-eff}
\begin{eqnarray}
H_{\rm eff} = \gamma_{xy} ( \sigma_1^x \sigma_2^x + \sigma_1^y
\sigma_2^y ) + \gamma_z \sigma_1^z \sigma_2^z + g_z ( \sigma_1^z +
\sigma_2^z) - i\tau g_{xy}^2 \left( 2 I_{12} + \sigma_1^x
\sigma_2^x + \sigma_1^y \sigma_2^y - \sigma_1^z - \sigma_2^z
\right).
\end{eqnarray}

\noindent Contribution of the identity operator $I_{12}$ in
$H_{\rm eff}$ affects only the probability of observing the
desired sequence of outcomes, but it does not affect physical
evolution of the normalized density operator
$\varrho_S(t)$, cf. Eq.~\eqref{normalized}. Therefore,
the physical dynamics of $\varrho_S(t)$ is governed by
the non-Hermitian Hamiltonian $H_{\rm eff}' = H_{\rm eff} + 2 i
\tau g_{xy}^2 I_{12}$, which reads as follows in the conventional
eigenbasis $\ket{00},\ket{01},\ket{10},\ket{11}$ of operator
$\sigma_1^z \sigma_2^z$:
\begin{eqnarray}\label{2 qubit Eff NHH}
H_{\rm eff}' = \left(%
\begin{array}{cccc}
  \gamma_z + 2 g_z + i 2 \tau g_{xy}^2 & 0 & 0 & 0 \\
  0 & -\gamma_z & 2 \gamma_{xy} - i 2 \tau g_{xy}^2 & 0 \\
  0 & 2 \gamma_{xy} - i 2 \tau g_{xy}^2 & -\gamma_z & 0 \\
  0 & 0 & 0 & \gamma_z - 2 g_z - i 2 \tau g_{xy}^2 \\
\end{array}%
\right).
\end{eqnarray}

\noindent The obtained Hamiltonian provides an adequate
description within the stroboscopic approximation, which is
justified if $g_{xy}\tau \ll
1$~\cite{rau-1963,giovannetti-2012,filippov-2017,fsp-2020}.
However, the quantity $g_{xy}^2 \tau$ can be comparable with
$\gamma_{xy}$ if the coupling strength $g_{xy} \gg \gamma_{xy}$.
If this is the case, the anti-Hermitian part of $H_{\rm eff}'$
cannot be neglected and should be properly taken into account.

In general scenario, the measurement repetition rate $\tau^{-1}$
can be time-dependent, i.e., the duration $\tau = \tau(t)$ in
between the sequential measurements can gradually vary with time
$t$ on a long timescale ($t \gg \tau(t)$). This leads to a
time-dependent Hamiltonian $H_{\rm eff}'(t)$.

\subsection{Two-qubit entanglement generation}

The time evolution operator $\mathcal{U}(t)$ of the two-qubit effective time-independent Hamiltonian in Eq.\eqref{2 qubit Eff NHH} can be easily derived.
It possesses the same structure of the Hamiltonian and turns out to be precisely
\begin{equation} \label{u}
\mathcal{U}(t)=
\begin{pmatrix}
^{-i(\gamma_z+2g_z)t/\hbar} ~ e^{2\tau g_{xy}^2t/\hbar} & 0 & 0 & 0 \\
0 & \cos\alpha & -i\sin\alpha & 0 \\
0 & -i\sin\alpha & \cos\alpha & 0 \\
0 & 0 & 0 & e^{-i(\gamma_z-2g_z)t/\hbar} ~ e^{-2\tau g_{xy}^2t/\hbar},
\end{pmatrix}.
\end{equation}
with $\alpha=2(\gamma_{xy}-i\tau g_{xy}^2)t\equiv(\gamma-i g)t$.

Since the dynamics of the two states $\ket{00}$ and $\ket{11}$ is
trivial, we concentrate on the dynamics within the dynamically
invariant Hilbert subspace spanned by $\ket{01}$ and $\ket{10}$
and governed by the $2 \times 2$ block. If the two qubits are
initially prepared in the pure state $\rho_S(0)=\ket{01}\bra{01}$,
following the scheme outlined in Secs. \ref{sec:nhrho} and
\ref{sec:filippov}, we get
\begin{equation}
\varrho_S^c(t)={\rho_S^c(t) \over \text{Tr}\{\rho_S^c(t)\}}={\mathcal{U}(t)\rho_S(0)\mathcal{U}^\dagger(t) \over \text{Tr}\{\mathcal{U}(t)\rho_S(0)\mathcal{U}^\dagger(t)\}}=
{1 \over |\cos(\alpha)|^2 + |\sin(\alpha)|^2}
\begin{pmatrix}
0 & 0 & 0 & 0 \\
0 & |\cos\alpha|^2 & i\cos\alpha\sin\alpha^* & 0 \\
0 & -i\cos\alpha^*\sin\alpha & |\sin\alpha|^2 & 0 \\
0 & 0 & 0 & 0
\end{pmatrix}.
\end{equation}
The (normalized) transition probability towards the state $\ket{10}$, $P_{01}^{10}$,  is then
\begin{equation}\label{Tr Prob}
P_{01}^{10}={|\sin\alpha|^2 \over |\cos(\alpha)|^2 + |\sin(\alpha)|^2}={\cos^2(\gamma t)\sinh^2(gt) + \sin^2(\gamma t)\cosh^2(gt) \over \cosh^2(gt) + \sinh^2(gt)}.
\end{equation}

\begin{figure}[htp]
\begin{center}
\subfloat[][]{\includegraphics[width=0.4\textwidth]{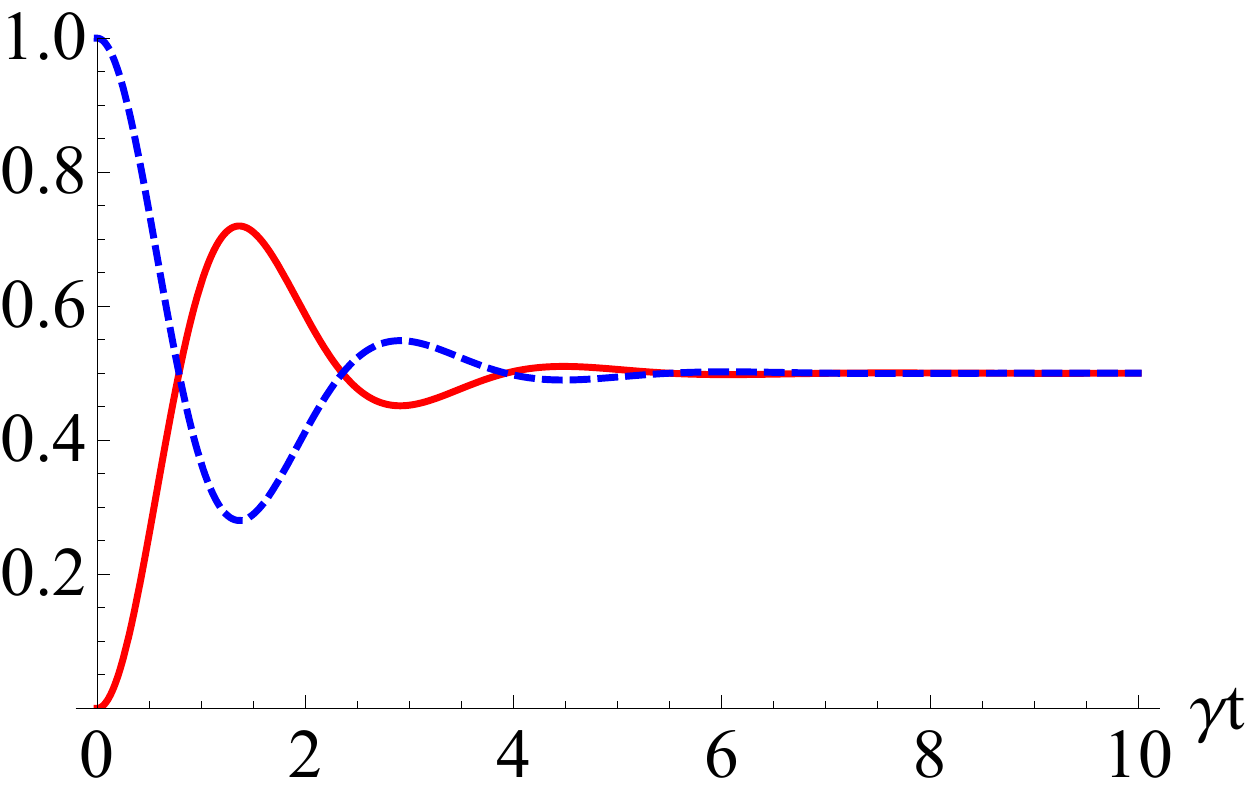}\label{fig:Pops}}
\qquad
\subfloat[][]{\includegraphics[width=0.4\textwidth]{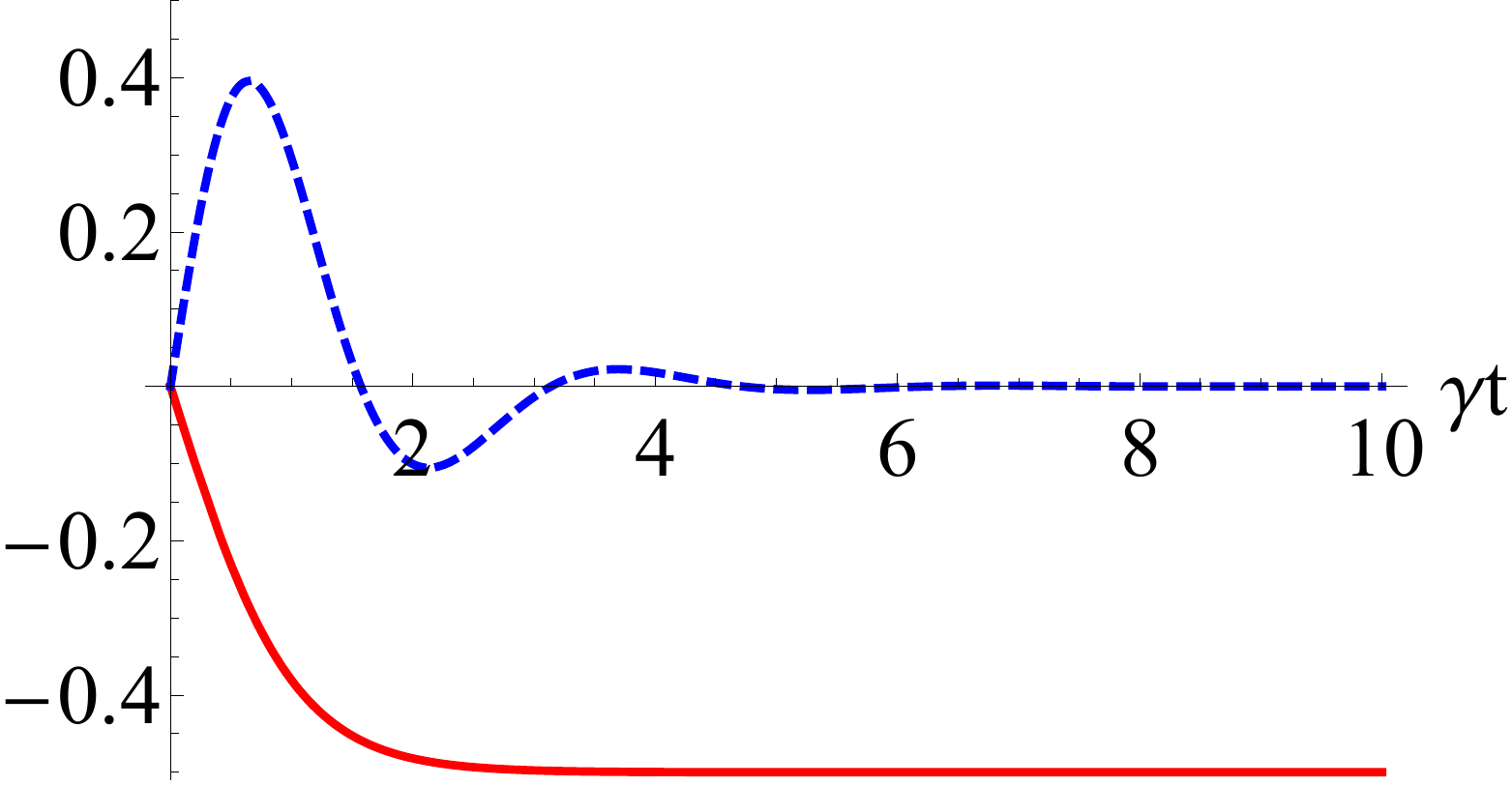}\label{fig:ReIm}}
\captionsetup{justification=raggedright,format=plain,skip=4pt}%
\caption{(Color online) a) Populations of the sates $\ket{10}$ (red solid line) and $\ket{01}$ (blue dashed line) when the two-qubit system is initially prepared in $\ket{01}$ for $\gamma=2g$; b) Real (solid red line) and imaginary (dashed blue line) part of the coherence $\bra{01}\varrho_S^c(t)\ket{10}$.}
\end{center}
\end{figure}

In Fig. \ref{fig:Pops} the two normalized populations are reported
in terms of the dimensionless parameter $\gamma t$ and for
$\gamma=2g$. The solid red (dashed blue) curve represents the
population of the state $\ket{10}$ ($\ket{01}$) and then the
transition probability in Eq. \eqref{Tr Prob}. We see that both
populations reach the value 1/2 at large times. In Fig.
\ref{fig:ReIm}, instead, we can see the time behaviours of the
real (solid red line) and imaginary (dashed blue line) parts of
the (normalized) coherence: $\bra{01}\varrho_S^c(t)\ket{10}$. We
notice that the coherence does not asymptotically vanish and,
rather, it becomes real and equal to $-1/2$, as it can be easily
verified by its analytical expression
\begin{equation}
\bra{01}\varrho_S^c(t)\ket{10}(t)=-{1 \over 2}{\sinh(2gt) - i \sin(2\gamma t) \over \cosh^2(gt) + \sinh^2(gt)}.
\end{equation}
The normalized asymptotic state reached by the two-qubit system turns out to be thus
\begin{equation}
\varrho_S^c(t\rightarrow\infty)={1 \over 2}
\begin{pmatrix}
0 & 0 & 0 & 0 \\
0 & 1 & -1 & 0 \\
0 & -1 & 1 & 0 \\
0 & 0 & 0 & 0
\end{pmatrix}
=\ket{\Psi^-}\bra{\Psi^-}, \qquad \ket{\Psi^-}={\ket{01}-\ket{10} \over \sqrt{2}},
\end{equation}
which is one of the the well known maximally entangled Bell
states. This result is in accordance with the fact that the
generalized von Neumann--Liouville equation \eqref{eq:drhodt}
preserves the purity of initial pure states.

In other words, when the system is initially prepared in a pure state,
the evolved state, according to Eq. \eqref{eq:drhodt}, remains a pure state
\cite{as-kz-2013,kz-2015}.
Conversely, mixed states change their purity during the time evolution \cite{as-kz-2013,kz-2015}.
Therefore, we have shown that, by initializing the two qubits in a pure separable state, we can generate an asymptotic pure entangled state through the procedure described in Sec. \ref{sec:filippov} based on repeated measurements on the third ancilla qubit.

\subsection{Effects of Hamiltonian anisotropy}

The same physical effect of entanglement generation for two
qubits, induced by repeated measurements on the third ancilla
qubit, does not occur if the system is initialized in either
$\ket{00}$ or $\ket{11}$. This fact is immediately clear from the
matrix form of the effective non-Hermitian Hamiltonian [Eq.
\eqref{2 qubit Eff NHH}] governing the dynamics of the two coupled
qubits. However, it is reasonable to argue that an appropriate
generalization of the three-spin model can lead to the appearance
of off-diagonal elements `connecting' the two states under
consideration. In this way, in the subspace spanned by $\ket{00}$
and $\ket{11}$ we may have a dynamics similar to the one we
brought to light before.

To this end, let us consider the most general model of the three spins, namely
\begin{equation}\label{Gen Ham}
\tilde{H}=\gamma_x\sigma_1^x\sigma_2^x+\gamma_y\sigma_1^y\sigma_2^y+\gamma_z\sigma_1^z\sigma_2^z+
\alpha_x\sigma_1^x\sigma_3^x+\alpha_y\sigma_1^y\sigma_3^y+\alpha_z\sigma_1^z\sigma_3^z+
\beta_x\sigma_2^x\sigma_3^x+\beta_y\sigma_2^y\sigma_3^y+\beta_z\sigma_2^z\sigma_3^z.
\end{equation}
In this case, the effective non-Hermitian Hamiltonian describing the dynamics of the spins 1 and 2 when the repeated-measurement technique is applied on the third spin, turns out to be (up to terms proportional to the identity operator)
\begin{equation}
\tilde{H}_{\text{eff}}=(\alpha_z+i\tau\alpha_x\alpha_y)\sigma_1^z+(\beta_z+i\tau\beta_x\beta_y)\sigma_2^z+(\gamma_z-i\tau\alpha_z\beta_z)\sigma_1^z\sigma_2^z+(\gamma_x-i\tau\alpha_x\beta_x)\sigma_1^x\sigma_2^x+(\gamma_y-i\tau\alpha_y\beta_y)\sigma_1^y\sigma_2^y.
\end{equation}
It is possible to easily verify that this Hamiltonian presents two independent subdynamics: one involving the two states $\{\ket{00},\ket{11}\}$ and the other involving the two remaining states $\{\ket{01},\ket{10}\}$.
The existence of these two dynamically invariant subspaces can be traced back to the existence of the following constant of motion $\sigma_1^z\sigma_2^z$.
In each subspace, thus, the two-spin system effectively behaves like a two-level system and we can write a fictitious two-level Hamiltonian for each subdynamics.
The matrix representation of the two-level Hamiltonian ruling the two-spin dynamics within the subspace spanned by $\{\ket{00},\ket{11}\}$ and $\{\ket{01},\ket{10}\}$ read respectively
\begin{equation}
\tilde{H}_{\text{eff}}^+=
\begin{pmatrix}
\gamma_z+\alpha_z+\beta_z+i\tau(\alpha_x\alpha_y+\beta_x\beta_y-\alpha_z\beta_z) & (\gamma_x-\gamma_y)-i\tau(\alpha_x\beta_x-\alpha_y\beta_y) \\
(\gamma_x-\gamma_y)-i\tau(\alpha_x\beta_x-\alpha_y\beta_y) & -[-\gamma_z+\alpha_z+\beta_z+i\tau(\alpha_x\alpha_y+\beta_x\beta_y+\alpha_z\beta_z)]
\end{pmatrix},
\end{equation}

\begin{equation}
\tilde{H}_{\text{eff}}^-=
\begin{pmatrix}
-\gamma_z+\alpha_z-\beta_z+i\tau(\alpha_x\alpha_y-\beta_x\beta_y+\alpha_z\beta_z) & (\gamma_x+\gamma_y)-i\tau(\alpha_x\beta_x+\alpha_y\beta_y) \\
(\gamma_x+\gamma_y)-i\tau(\alpha_x\beta_x+\alpha_y\beta_y) & -[\gamma_z+\alpha_z-\beta_z+i\tau(\alpha_x\alpha_y-\beta_x\beta_y-\alpha_z\beta_z)]
\end{pmatrix},
\end{equation}
where the superscripts $+$ and $-$ refer to the two values $\pm 1$ of the constant of motion $\sigma_1^z\sigma_2^z$.
The operatorial form of $\tilde{H}_{\text{eff}}^\pm$ in terms of dynamical variable of a fictitious spin-1/2, omitting terms with no influence in the two-qubit dynamics, reads
\begin{equation}
\begin{aligned}
\tilde{H}_{\text{eff}}^{\pm}&=\Omega_\pm\sigma^z+\omega_\pm\sigma^x, \\
\Omega_{\pm}&=\alpha_z\pm\beta_z+i\tau(\alpha_x\alpha_y\pm\beta_x\beta_y)\equiv \mu_z+i\nu_z, \\
\omega_{\pm}&=(\gamma_x\mp\gamma_y)-i\tau(\alpha_x\beta_x\mp\alpha_y\beta_y)\equiv
\mu_x+i\nu_x.
\end{aligned}
\end{equation}
We get the model previously analysed by putting
$\gamma_x=\gamma_y$, $\alpha_x=\alpha_y=\beta_x=\beta_y$, and
$\alpha_z=\beta_z$. We see, in fact, that the first two conditions
make the off-diagonal entries in $\tilde{H}_{\text{eff}}^+$ equal
zero, as expected.

The time evolution operator related to $\tilde{H}_{\text{eff}}^+$,
that is, restricted to the subspace spanned by $\ket{00}$ and
$\ket{11}$, turns out to be
\begin{equation}\label{u tilde +}
\tilde{u}^+(t)=
\begin{pmatrix}
\cos(\nu t)-i{\Omega_+ \over \nu}\sin(\nu t) & -i{\omega_+ \over \nu}\sin(\nu t) \\
-i{\omega_+ \over \nu}\sin(\nu t) & \cos(\nu t)+i{\Omega_+ \over
\nu}\sin(\nu t)
\end{pmatrix},
\qquad \nu= \sqrt{\Omega_+^2+\omega_+^2}.
\end{equation}
We note that for $\Omega_+=0$ we get the analogous form of the
time evolution operator in Eq. \eqref{u}.

\begin{figure}[htp]
\begin{center}
\subfloat[][]{\includegraphics[width=0.4\textwidth]{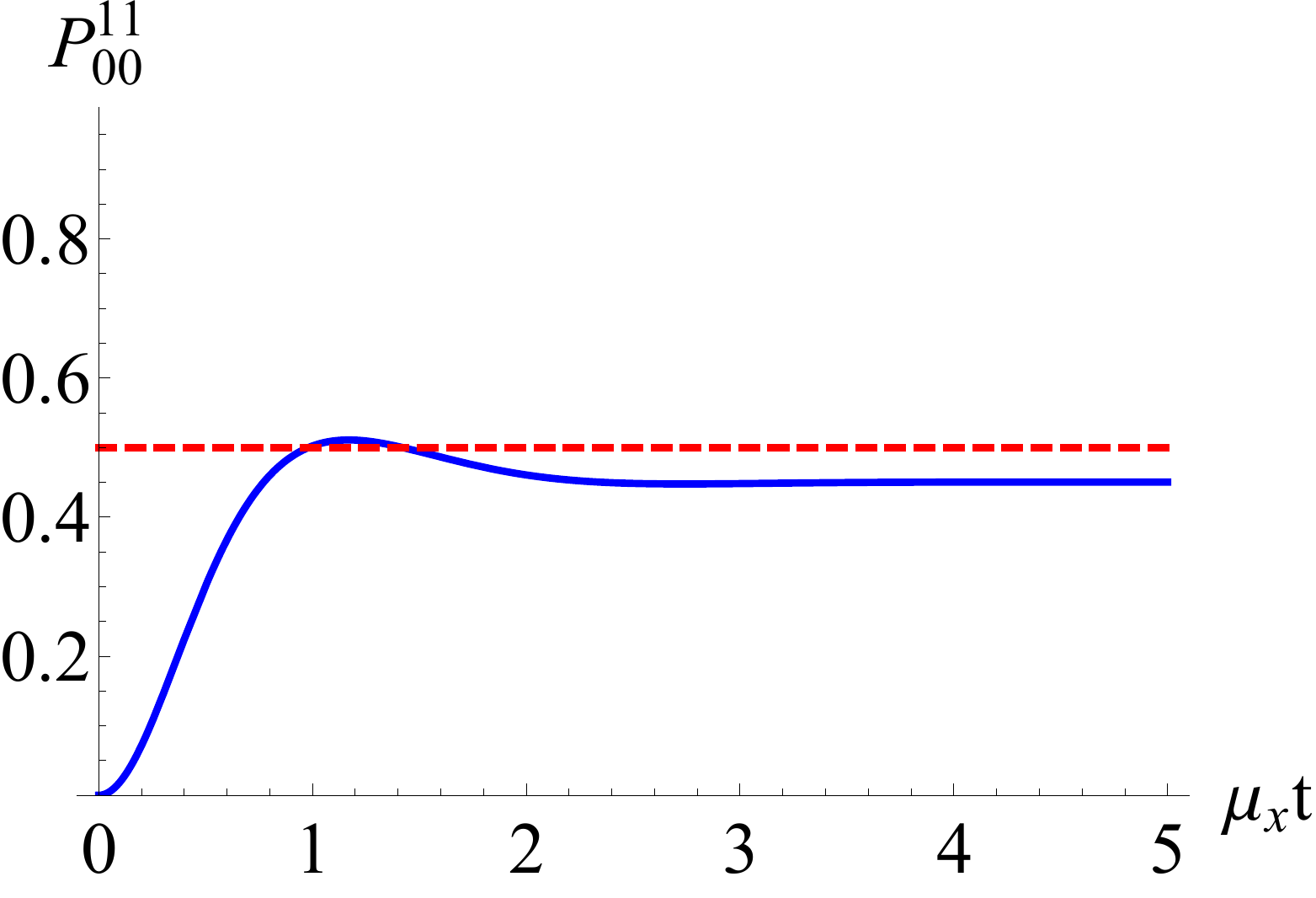}\label{fig:ri10RI}}
\qquad
\subfloat[][]{\includegraphics[width=0.4\textwidth]{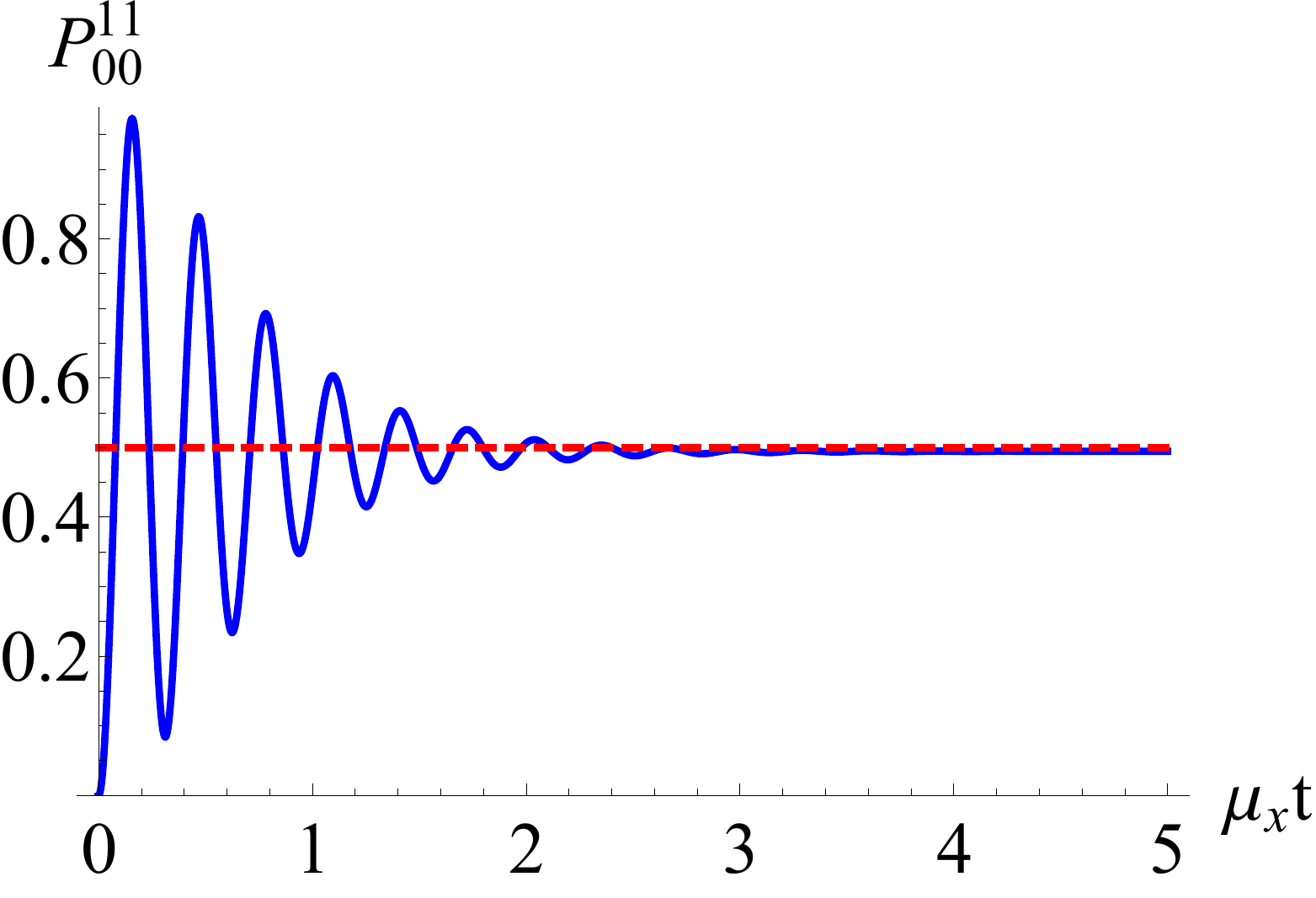}\label{fig:r10i100RI}}
\qquad
\subfloat[][]{\includegraphics[width=0.4\textwidth]{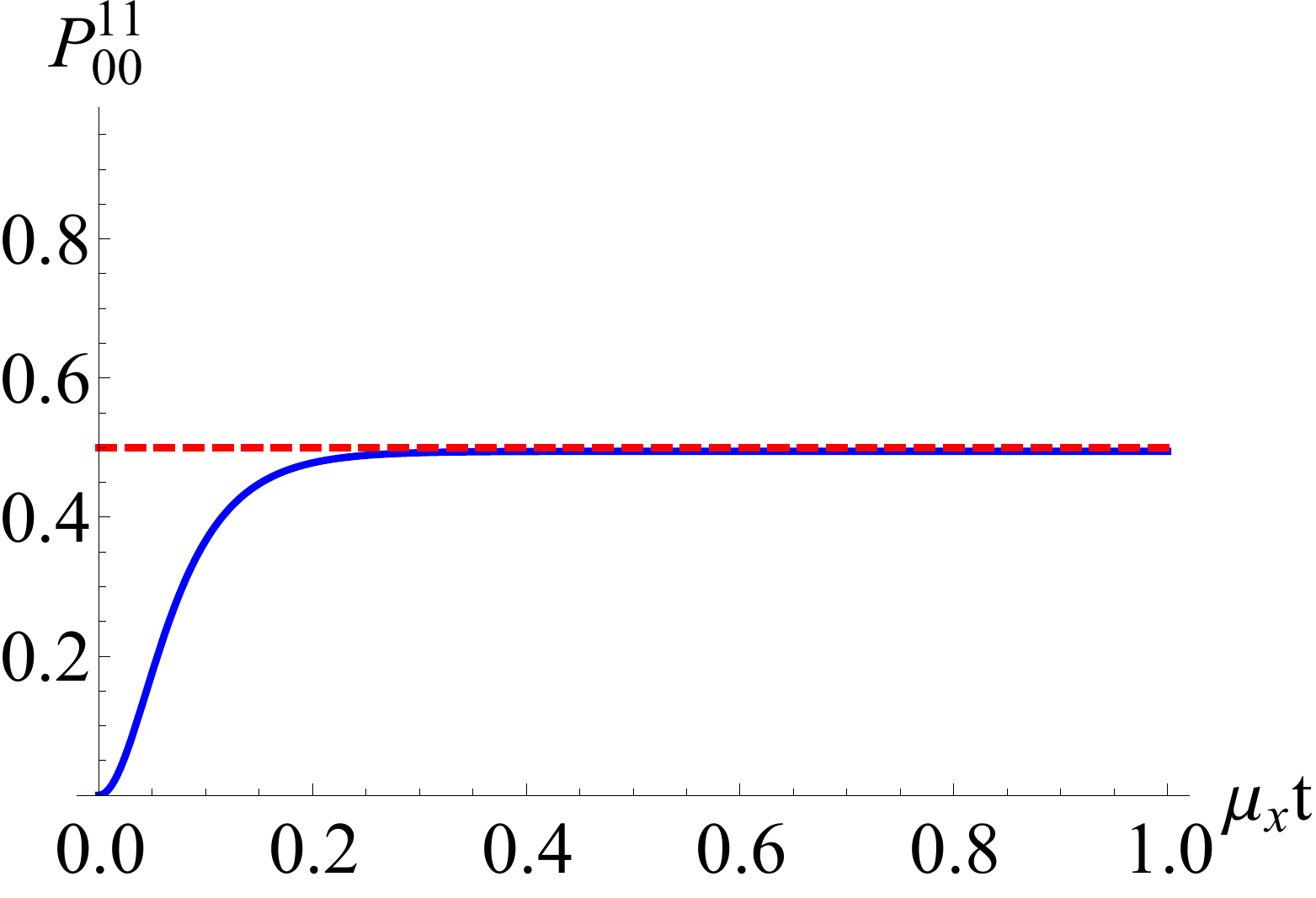}\label{fig:i10r100RI}}
\qquad
\subfloat[][]{\includegraphics[width=0.4\textwidth]{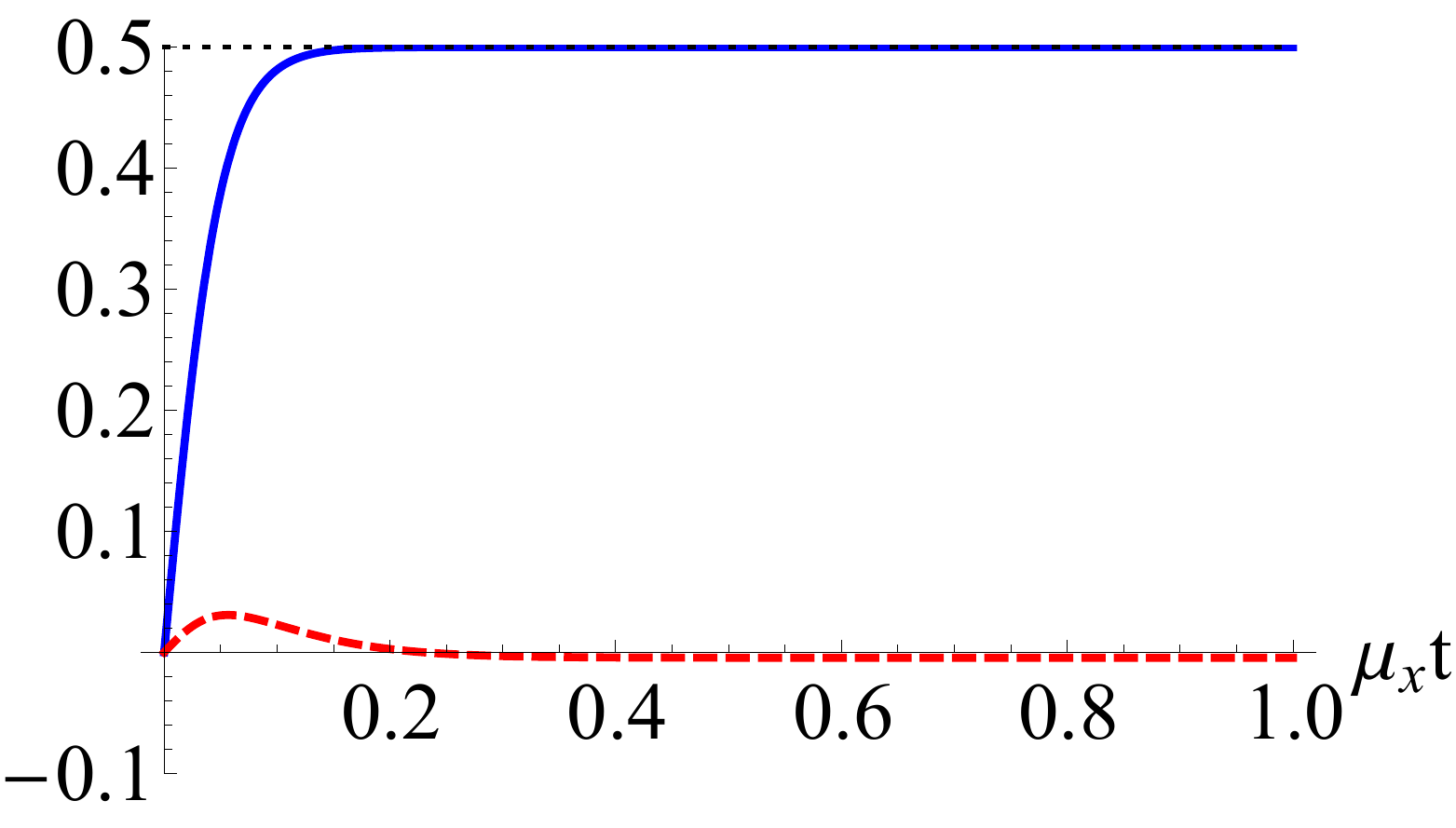}\label{fig:ReIm2}}
\captionsetup{justification=raggedright,format=plain,skip=4pt}%
\caption{(Color online) a) Populations of the sates $\ket{11}$ (solid blue line) when the two-qubit system is initially prepared in $\ket{00}$ for a) $\mu_x=\nu_y=10\mu_z=10\nu_z$, b) $\mu_x=10\nu_y=100\mu_z=100\nu_z$, c) $10\mu_x=\nu_y=100\mu_z=100\nu_z$ (the dashed red line represents $P_{00}^{11}=1/2$); d) Real (solid blue line) and imaginary (dashed red line) part of the coherence $\bra{01}\tilde{\varrho}_S^c(t)\ket{10}$ when $10\mu_x=\nu_y=100\mu_z=100\nu_z$. Plots are reported versus the dimensionless parameter $\mu_x~t$.}
\end{center}
\end{figure}

In Figs. \ref{fig:ri10RI}, \ref{fig:r10i100RI}, and
\ref{fig:i10r100RI} the population of the state $\ket{11}$ is
reported when the two-spin system is initially prepared in
$\tilde{\rho}_S(0)=\ket{00}\bra{00}$. We can qualitatively
appreciate that different relative weights of the parameters
$\mu_x$, $\nu_x$, $\mu_z$, and $\nu_z$ give rise to different time
behaviours. In all three cases we chose the favourable condition
$\Omega_+ \ll \omega_+$ to generate an asymptotic entangled state.
From Fig. \ref{fig:ReIm2}, in fact, we see that for
$10\mu_x=\nu_y=100\mu_z=100\nu_z$ (Fig. \ref{fig:i10r100RI}), the
coherence of the state
$\tilde{\varrho}_S^c(t)=\tilde{\rho}_S^c(t)/\text{Tr}\{\tilde{\rho}_S^c(t)\}$
becomes real at large times, meaning that the two-spin system
asymptotically reaches the state
\begin{equation}
\tilde{\varrho}_S^c(t \rightarrow \infty)={1 \over 2}
\begin{pmatrix}
1 & 0 & 0 & 1 \\
0 & 0 & 0 & 0 \\
0 & 0 & 0 & 0 \\
1 & 0 & 0 & 1
\end{pmatrix}
=\ket{\Phi^+}\bra{\Phi^+}, \qquad \ket{\Phi^+}={\ket{00}+\ket{11} \over \sqrt{2}}.
\end{equation}

This result shows that, under the generalized model in Eq.
\eqref{Gen Ham} and the repeated-measurement procedure, it is
possible to generate maximally entangled state in the subspace
spanned by $\ket{00}$ and $\ket{11}$ too. The presence of
anisotropy in the exchange interaction between the two spins under
consideration and/or between each spin with the ancilla, in fact,
makes the half-transition $\ket{11} \leftrightarrow \ket{00}$
possible, producing, thus, detectable physical effects which would
be absent under the more isotropic model in Eq.
\eqref{three-spin-Hamiltonian}. Therefore, it means that by
studying the dynamics in this subspace we can get information
about the level of (an)isotropy of the coupling existing between
the two spins and between each spin with the ancilla.

\section{Conclusive Remarks}

Reference \cite{luchnikov-2017} reports an original experimental protocol implementing the quantum dynamics of a finite-dimensional system $S$ generated by a non-Hermitian Hamiltonian operator. In accordance with this scheme, firstly $S$ is appropriately coupled with a finite-dimensional quantum ancilla subsystem $A$ and then the time evolution of $S$, conditioned by a Zeno measurement protocol applied on $A$ only, is observed at any intermediate step. In accordance with Ref. \cite{luchnikov-2017}, the reduced density matrix of $S$, stemming from the progression of collapses induced in this way on the state of the combined system $S+A$, evolves under the action of an
effective non-Hermitian Hamiltonian which may be explicitly constructed in the so called stroboscopic regime limit.
In the present work, this method was applied to a system $S$ composed by a two-qubit system
interacting with a third ancilla qubit. The scope is to demonstrate the effectiveness and usefulness of the protocol to predict the quantum dynamics of the pair of qubits conditioned by a
quantum Zeno measurement protocol applied to the ancilla only.

First, we took into account pairwise Heisenberg interactions
between the three spins so that the two relevant spins (system)
are identically coupled to the ancilla qubit (the case of
reflectional symmetry). The method proposed in Ref.
\cite{luchnikov-2017} proved to be successful leading us to an
effective non-Hermitian time-independent two qubit model. The
exact solution of the dynamical problem has been simplified by
analysing the different dynamically invariant subspaces related to
the symmetry possessed by the effective Hamiltonian. This
symmetry-based approach turned out to be useful to study and solve
dynamical problems related to more complex interacting spin
systems subjected to time-dependent fields
\cite{GMN,GMIV,GBNM,GLSM,GMGIM,GVMqubits,GVMqutrits,GMMM}. By
focusing our attention on the sub-dynamics involving the two-qubit
states $\ket{10}$ and $\ket{01}$, we brought to light the
possibility of generating maximally entangled states of the two
qubits. Therefore, we showed that the technique based on repeated
measurements on the ancilla qubit can induce quantum correlations
on the two-qubit subsystem.

A second interesting aspect consists in the detectable physical effects on the dynamics of the two-qubit system stemming from the isotropy level of the spin interactions.
We know that the type of interaction could considerably affect the system dynamics giving rise to remarkable physical effects \cite{Wang,LiuKong,Liu,GNMV}.
We generalized the model by analysing anisotropic Heisenberg interactions between the three spins.
In this case, of course, the effective non-Hermitian two-qubit Hamiltonian turned out to be more complicated.
However, conserved symmetries possessed by the Hamiltonian ensured again the existence of two dynamically invariant sub-dynamics making simpler the study and solution of the two-qubit dynamical problem.
We demonstrated that the anisotropic interactions can generate transitions in the subspace involving the two-qubit states $\ket{00}$ and $\ket{11}$ which were hindered, instead, in the isotropic scenario.
So, the possibility of generating maximally entangled states in both sub-dynamics is a transparent and relatively experimentally easy way both to manifest and to get information about the isotropy level of the qubit interactions.

A further important result achieved in this paper is that described in Section \ref{sec:inverse}. It may be described as the inverse of the the protocol reported in Ref. \cite{luchnikov-2017}. In fact, starting from a non-Hermitian Hamiltonian model at will for an arbitrary system $S$, it introduces an easy and universal recipe to construct an Hermitian Hamiltonian model for the system $S+A$ where $A$ is a qubit system.
The importance of this original inverse protocol stems from the fact that it holds whatever the system $S$ and its non-Hermitian prescribed model is. When $S$ is finite-dimensional the application of the direct protocol leads to the assigned non-Hermitian Hamiltonian. Thus, for example, we may start from a pseudo-Hermitian Hamiltonian describing a finite system $S$, to generate a physical scenario where the time behaviour of $S$ may be well simulated under 
stroboscopic conditions established in Section \ref{sec:filippov}.

It is interesting to point out that theoretical investigations on non-Hermitian Hamiltonians find useful applications not only in the quantum realm but also in the classical one.
Let us think about the non linear optics branch \cite{Boyd}, for example.
A first optical scenario deserving to be mentioned is the one regarding the laser-induced continuum structure problem \cite{Knight} which has been deeply investigated \cite{VitanovLICS} and experimentally confirmed \cite{Halfmann}.
More recently, instead, a lot of attention has been paid to interacting waveguides.
It is possible to show that, under appropriate physical conditions, the dynamics of these systems can be well described by a Schr\"odinger-like equation where the spatial variable plays the role of time in the standard Schr\"odinger equation \cite{Longhi}.
It is worth that experimentalists, through the appropriate choice of materials and laser-based techniques, are able to control some parameters in such a way that the Hamiltonian ruling the dynamics turns out to be non-Hermitian \cite{RuterGuo}.
Therefore, at the light of these examples too, we understand how many intriguing aspects about the dynamics of both quantum and classical physical systems may still need to be found.

A future perspective of the present work could be to investigate
the cases in which the parameter $\tau$ or the total Hamiltonian
for $S+A$ are considered to be time-dependent, taking into account
exactly solvable non-Hermitian scenarios recently proposed
\cite{grimaudo-2018,GdCNM}. Moreover, one can concentrate in the
application of the theoretical method of Ref.
\cite{luchnikov-2017} in more complex cases, like two-qubit
systems immersed in a quantum oscillator environment. In this
case, a fruitful comparison with other approaches
\cite{Kapral,SHGM,SGHM} developed to face with this kind of
problem is possible.

\acknowledgments{The work of S.N. Filippov was performed at the
Steklov International Mathematical Center and supported by the
Ministry of Science and Higher Education of the Russian Federation
(agreement no. 075-15-2019-1614).}

\end{document}